\documentclass[twocolumn,showpacs,preprintnumbers,amsmath,amssymb]{revtex4}

\usepackage{graphicx}
\usepackage{dcolumn}
\usepackage{bm}

\begin{document}

\preprint{}

\title{Necessary and sufficient conditions for local creation of quantum correlation}

\author{Xueyuan Hu}
\email{xyhu@iphy.ac.cn}
\author{Heng Fan}
\author{D. L. Zhou}
\author{Wu-Ming Liu}
\affiliation{Beijing National Laboratory for Condensed Matter
Physics, Institute of Physics, Chinese Academy of Sciences, Beijing
100190, China}
\date{\today}

\begin{abstract}
Quantum correlation can be created by local operation from some
initially classical states. We prove that the necessary and
sufficient condition for a local trace-preserving channel to create
quantum correlation is that it is not a commutativity-preserving
channel. This condition is valid for arbitrary finite dimension
systems. We also derive the explicit form of
commutativity-preserving channels. For a qubit, a
commutativity-preserving channel is either a completely decohering
channel or a mixing channel. For a three-dimension system (qutrit),
a commutativity-preserving channel is either a completely decohering
channel or an isotropic channel.
\end{abstract}

\pacs{03.65.Ud, 03.65.Yz, 03.67.Mn}
\maketitle

Quantum correlation is the unique phenomenon of quantum physics and
believed to be a resource for quantum information processes which
can generally surpass the corresponding classical schemes. Many
previous studies focus on entanglement, a well-known quantum
correlation, since its apparent role in teleportation, superdense
coding \cite{tele,PhysRevA.54.1869}, etc. Recently, measures of the
nonclassicalness of correlation, such as quantum discord
\cite{PhysRevLett.88.017901} and quantum deficit
\cite{PhysRevLett.89.180402,PhysRevA.71.062307}, began to attract
much attention since the discovery that some quantum information
schemes can be realized without entanglement but with a positive
quantum discord
\cite{PhysRevLett.100.050502,PhysRevLett.107.080401}. Much progress
has been made to quantify the amount of quantum correlation in
different physical systems
\cite{PhysRevA.77.042303,PhysRevA.81.042105} and to give it
intuitive and operational interpretations. It is shown that quantum
discord can be operationally interpreted as the minimum information
missing from the environment \cite{arXiv:1110.1664v1}. One-way
quantum deficit \cite{RevModPhys.81.1,PhysRevA.67.012320} has been
found as the reason for entanglement irreversibility
\cite{PhysRevLett.107.020502} and can be related to quantum
entanglement via an interesting scheme
\cite{PhysRevLett.106.160401,PhysRevLett.106.220403}.

Quantum noise usually plays a destructive role in quantum
information process. However, there are situations that local
quantum noise can enhance nonlocal quantum properties for some mixed
quantum states. For example, local amplitude damping can increase
the average teleportation fidelity for a class of entangled states
\cite{PhysRevA.62.012311,PhysRevA.78.022334,PhysRevA.81.054302}.
Quantum discord can also be increased or created by local noise
\cite{PhysRevA.84.022113,PhysRevA.85.010102,PhysRevA.85.022108}. An
interesting result is that any separable state with positive quantum
discord can be produced by local positive operator-valued measure
(POVM) on a classical state in a larger Hilbert space
\cite{PhysRevA.78.024303}. In fact, almost all states in the Hilbert
space contains quantum correlation, and an arbitrary small
disturbance can drive a classical state into a quantum state with
nonzero quantum correlation \cite{PhysRevA.81.052318}.
Counterintuitively, it has recently been discovered that mixedness
is as important as entanglement for quantum correlation. In
particular, some mixed states contain more quantum discord than that
of maximally entangled pure state when the dimension of the system
is large enough \cite{PhysRevLett.106.220403}. Thus it is of
interest to know how is the effect of mixedness on the quantum
correlation of quantum states. The condition for local increase of
quantum correlation has been derived for the qubit case
\cite{PhysRevLett.107.170502}, and it has been pointed out that this
condition is not valid for high-dimension systems.

In this article, we derive a simple necessary and sufficient
condition for a local channel to create quantum correlation in some
half-classical states, which is valid for arbitrary finite dimension
systems. A trace-preserving local channel can create quantum
correlation if and only if it is not a commutativity-preserving
channel. For qubit case, we show that a commutativity-preserving
channel is either a mixing channel or a completely decohering
channel. This confirms the result in Ref.
\cite{PhysRevLett.107.170502}. For the qutrit case, quantum
correlation can be created by a local channel in some half-classical
input states if and only if the channel is neither a completely
decohering channel nor an isotropic channel. We also analyze the
reason for a local mixing channel to create quantum correlation in
qutrit situation and then give a conjecture to extend the result of
qutrits to arbitrary finite dimension systems.

The total correlation between two quantum systems is composed of
classical and quantum correlations. From this point of view, quantum
correlation is defined as the difference between total and classical
correlations. Therefore, various measures of quantum correlation
defined on one party of a composite system vanish for exact the same
class of states, which is called half-classical states. Because
classical correlation is defined by the correlation that can be
revealed by local measurements, a state $\rho_{AB}$ is
half-classical on $B$ if and only if there exist a measurement on
$B$ that does not affect the total state. As proved in Ref.
\cite{arXiv:0807.4490v1} a half-classical state on $B$ can be
written as
\begin{equation}
\rho_{AB}=\sum_ip_i\rho_A^{\alpha_i}\otimes|\alpha_i\rangle_B\langle\alpha_i|.\label{half_classical}
\end{equation}
where $\{|\alpha_i\rangle_B\}$ consist an orthogonal basis for the
Hilbert space of subsystem $B$, and $\rho_A^{\alpha_i}$ are
corresponding density matrices of $A$. The subsystem $A$ can be a
single quantum particle or an ensemble of quantum particles. In the
following, by quantum correlation, we mean quantum correlation
defined on subsystem $B$. The main purpose of this paper is to
characterize the channel $\Lambda_B$ satisfying
\begin{equation}
\mathrm I_A\otimes\Lambda_B(\rho_{AB})\in\mathcal
D_0,\forall\rho_{AB}\in\mathcal D_0,
\end{equation}
where $\mathcal D_0$ is the set of half-classical states. Before
providing the condition, we first introduce a class of quantum
channels, which we call commutativity-preserving channels.

Definition 1 (commutativity-preserving channel): a
commutativity-preserving channel $\Lambda^{\mathrm{CP}}$ is the
channel that can preserve the commutativity of any input density
operators, i.e.,
\begin{equation}
\label{criterion2}[\Lambda^{\mathrm{CP}}(\xi),\Lambda^{\mathrm{CP}}(\xi')]=0
\end{equation}
holds for any density operators satisfying $[\xi,\xi']=0$.

It is worth mentioning an equivalent definition of a
commutativity-preserving channel. A channel $\Lambda$ is a
commutativity-preserving channel if and only if
\begin{equation}
\label{c3}[\Lambda(\phi),\Lambda(\psi)]=0
\end{equation}
holds for any pure states satisfying $\langle\phi|\psi\rangle=0$.
The ``only if'' part is obtained directly by choosing
$\xi=|\phi\rangle\langle\phi|$ and $\xi'=|\psi\rangle\langle\psi|$.
Conversely, if Eq. (\ref{c3}) holds, by writing $\xi$ and $\xi'$ on
their common eigenbasis, we arrive at Eq. (\ref{criterion2}).

Now we are ready to prove the first main result of this paper. It
holds for arbitrary finite-dimension systems.

Theorem 1: A channel $\Lambda$ acting on subsystem $B$ can create
quantum correlation between subsystems $A$ and $B$ for some input
half-classical state $\rho_{AB}$ if and only if it is not a
commutativity-preserving channel.

Proof: Any separable state can be written as
\begin{equation}
\xi_{AB}=\sum_ip_i\xi_i^A\otimes\xi_i^B,\label{separable}
\end{equation}
where $\xi_i^A$ are linearly independent. We will first prove that
$\xi_{AB}$ is a half-classical state if and only if
\begin{equation}
[\xi_i^B,\xi_j^B]=0,\ \forall i,j.\label{criterion1}
\end{equation}
For proving the ``only if'' part, we notice that for any
half-classical state, there exist a measurement basis
$\Pi_B^{\alpha_i}$ that does not affect the state. Therefore,
\begin{equation}
\sum_ip_i\xi_i^A\otimes(\xi_i^B-\Pi_B^{\alpha_j}\xi_i^B\Pi_B^{\alpha_j})=0.
\end{equation}
Because $\xi_i^A$ are linearly independent, $\xi_i^B$ is diagonal on
$\{\Pi_B^{\alpha_j}\}$ and thus satisfies Eq. (\ref{criterion1}).
Conversely, if Eq. (\ref{criterion1}) holds, $\xi^B_i$ and $\xi^B_j$
share common eigenvectors for any $i$ and $j$. By choosing these
eigenvectors as the basis for von Neumann measurement, the state
does not change after the measurement, which means that $\xi_{AB}$
is a half-classical state. Now consider an arbitrary half-classical
state in form of Eqs. (\ref{separable}) and (\ref{criterion1}) as
the input state, the channel $\Lambda$ acting on subsystem $B$ leads
the state to $\xi'_{AB}\equiv\mathrm
I_A\otimes\Lambda_B(\xi_{AB})=\sum_ip_i\xi_i^A\otimes\Lambda(\xi_i^B)$,
which is still a half-classical state if and only if
\begin{equation}
[\Lambda(\xi_i^B),\Lambda(\xi_j^B)]=0,
\end{equation}
for arbitrary choice of $\xi_i^B$ and $\xi_j^B$ satisfying Eq.
(\ref{criterion1}). This is just the definition of a
commutativity-preserving channel. Therefore, the channel $\Lambda$
can create quantum correlation for some input half-classical states
if and only if it is not a commutativity-preserving channel. This
completes the proof.

The rest of this paper is devoted to expose the exact form of a
commutativity-preserving channel.

Since $[\mathrm I,\rho]=0,\forall\rho$, we obtain a necessary
condition for a commutativity-preserving channel
\begin{equation}
[\Lambda(\mathrm I),\Lambda(\rho)]=0,\forall\rho.\label{c5}
\end{equation}
When $B$ is a qubit, Eq. (\ref{c5}) is also the sufficient
condition. The reason is as follows. By using the linearity of
$\Lambda$, the left hand side of Eq. (\ref{c3}) can be written as
\begin{eqnarray}
&&\frac12[\Lambda(|\phi\rangle\langle\phi|+|\psi\rangle\langle\psi|),\Lambda(|\psi\rangle\langle\psi|-|\phi\rangle\langle\phi|)]\nonumber\\
&=&\frac12[\Lambda(\mathrm
I),\Lambda(u\sigma^zu^{\dagger})],\label{c4}
\end{eqnarray}
where $|\psi\rangle=u|0\rangle$ and $|\phi\rangle=u|1\rangle$. Since
any qubit state $\rho$ can be decomposed as $\rho=(\mathrm
I+n_x\sigma_x+n_y\sigma_y+n_z\sigma_z)/2$, Eq. (\ref{c3}) is
equivalent to Eq. (\ref{c5}). From this observation, we can see that
a qubit channel $\Lambda$ is commutativity-preserving if and only if
it is one of the following two cases:

Case 1: $\Lambda(\mathrm I)=\mathrm I$, which means that $\Lambda$
is a unital channel. Here we define a mixing channel
$\Lambda^{\mathrm M}$ as
\begin{equation}
S(\Lambda^{\mathrm M}(\rho_S))\geq S(\rho_S),\ \forall
\rho_S,\label{mixing_channel_d}
\end{equation}
where $S(\rho)\equiv-\mathrm{Tr}(\rho\log_2\rho)$ is the von Neumann
entropy. It is worth mentioning that when a channel is a mixing
channel, its extension to larger systems $\mathrm
I_A\otimes\Lambda_S^{\mathrm M}$ is still a mixing channel. As
proved in Ref. \cite{RevModPhys.50.221}, a mixing channel is
equivalent to a unital channel.

Case 2: $\Lambda (\mathrm I) \neq \mathrm I$. Then the diagonal
basis of $\Lambda (\mathrm I)$ is specified. According to Eq. (8),
the two matrices $\Lambda(\rho)$ and $\Lambda(\mathrm I)$ share
common eigenvectors. In other words, the channel $\Lambda$ takes any
input state $\rho$ to a diagonal form on the eigenbasis of
$\Lambda(\mathrm I)$, and is thus a completely decohering channel.

Therefore, when $B$ is a qubit, a commutativity-preserving channel
is either a mixing channel or a completely decohering channel. This
confirms the result in Ref. \cite{PhysRevLett.107.170502}.

In the following, we will move on to study the exact form of a
commutativity-preserving channel for high-dimension cases.

Definition 2 (isotropic channel) An isotropic channel is of the form
\begin{equation}
\Lambda^{\mathrm{iso}}(\rho)=p\Gamma(\rho)+(1-p)\frac{\mathrm I}{d},
\end{equation}
where $\Gamma$ is any linear channel that preserves the eigenvalues
of $\rho$. According to Ref. \cite{EnPC}, $\Gamma$ is either a
unitary operation or unitarily equivalent to transpose. Parameter
$p$ is chosen to make sure that $\Lambda$ is a completely positive
channel. In particular, $-1/(d-1)\leq p\leq1$ when $\Gamma$ is a
unitary operation, and $-1/(d-1)\leq p\leq1/(d+1)$ when $\Gamma$ is
unitarily equivalent to transpose.

Theorem 2: Consider the half-classical input state in Eq.
(\ref{half_classical}) with $B$ a qutrit, a channel $\Lambda$ can
not create quantum correlation in any half-classical input state if
and only if $\Lambda$ is either a completely dechering channel or an
isotropic channel.

Proof: Writing the eigen-decomposition of $\Lambda(\mathrm I)$ as
\begin{equation}
\Lambda(\mathrm I)=\sum_{i=1}^{N}\lambda_i\mathrm
I_{r_i}.\label{iout}
\end{equation}
Here $\sum_{i=1}^{N}r_i=\sum_{i=1}^{N}r_i\lambda_i=3$,
$\lambda_i\geq0$, $r_i$ are positive integers, and $\mathrm I_{r_i}$
are identities of the $r_i$-dimension subspace $\mathcal V_{r_i}$.
From Eq. (\ref{c5}) we have
\begin{equation}
\Lambda(\rho)=\sum_{i=1}^{N}q_i\xi^{\rho}_{r_i},\forall\rho.\label{crho}
\end{equation}
where $\xi^{\rho}_{r_i}$ is a density operator on $\mathcal
V_{r_i}$.

Clearly, when the eigenvectors of $\Lambda(\mathrm I)$ are
nondegenerate, i.e., $N=3$ and Eq. (\ref{iout}) becomes
$\Lambda(\mathrm I)=\sum_{i=1}^{3}\lambda_i\Pi_i$, the channel
$\Lambda$ is a completely decohering channel, since it takes any
input state $\rho$ to a diagonal form on basis $\{\Pi_i\}$. When two
or three eigenvectors of $\Lambda(\mathrm I)$ are degenerate, we
study the eigendecomposition of $\Lambda(\phi)$ for a pure input
state $|\phi\rangle$
\begin{equation}
\Lambda(\phi)=\sum_{i=1}^{N^{\phi}}\lambda^{\phi}_i\mathrm
I_{r_i(\phi)},\label{phiout}
\end{equation}
where $N^{\phi}\geq N$ and $\mathcal V_{r_i(\phi)}\subseteq\mathcal
V_{r_j}$. When none of $\Lambda(\phi)$ breaks the degeneracy of
eigenvectors of $\Lambda(\mathrm I)$, i.e., $N^{\phi}=N$ and
$\mathcal V_{r_i(\phi)}=\mathcal V_{r_i}$, the channel is also a
completely decohering channel. Now we focus on the case that some
$\Lambda(\phi)$ can break the degeneracy of eigenvectors of
$\Lambda(\mathrm I)$, i.e., $N^{\phi}>N$ and $\mathcal
V_{r_i(\phi)}\subset\mathcal V_{r_j}$ for some $i$. Let
$\{|\phi_k\rangle\}_{k=0}^2$ be a basis of the three-dimension
Hilbert space and $|\phi_0\rangle$ be the pure input state whose
corresponding output state $\Lambda(\phi_0)$ has the most different
eigenvalues. It means that $N^{\phi_0}\geq N^{\phi},\forall\phi$.

Case 1: For any state
$|\phi_0^{\bot}\rangle=c_1|\phi_1\rangle+c_2|\phi_2\rangle$ which is
orthogonal to $|\phi_0\rangle$, we have
$N^{\phi_0^{\bot}}=N^{\phi_0}$ and $\mathcal
V_{r_i(\phi_0^{\bot})}=\mathcal V_{r_i(\phi_0)}$.Then for arbitrary
input state $\varphi=\sum_{i=0}^2c_i|\phi_i\rangle$, we have
\begin{equation}
[\Lambda(\varphi),\Lambda(c_2^*|\phi_1\rangle-c_1^*|\phi_2\rangle)]=0.
\end{equation}
Therefore, $\Lambda(\varphi)$ is diagonal on the same basis as
$\Lambda(\phi_0)$.

Case 2: There exist a pure state, say $|\phi_2\rangle$, whose
corresponding output state $\Lambda(\phi_2)$ does not break as much
degeneracy as $\Lambda(\phi_0)$, i.e., $N^{\phi_2}<N^{\phi_0}$. We
will first prove that for any pure state
$|\varphi_{01}\rangle=|\phi_0\rangle-\beta_0|\phi_1\rangle$ in
2-dimension subspace $\mathcal W_{2}^{\phi_2}$, the output state is
diagonal on the same basis as $\Lambda(\phi_0)$, say $\{\Pi_i\}$. We
introduce
$|\varphi(\beta)\rangle=|\varphi_{01}\rangle+\beta|\phi_2\rangle$
and
$|\varphi_{02}(\beta)\rangle=\beta^*|\phi_0\rangle-|\phi_2\rangle$.
Notice that $\langle\varphi_{02}(\beta)|\varphi(\beta)\rangle=0$, we
have
\begin{equation}
[\Lambda(\varphi_{02}(\beta)),\Lambda(\varphi(\beta))]=0.\label{degenerate}
\end{equation}
Because $\sum_{k=0}^2|\phi_k\rangle\langle\phi_k|=\mathrm I$, we
have $N^{\phi_1}=N^{\phi_0}$ and $\mathcal V_{r_i(\phi_1)}=\mathcal
V_{r_i(\phi_0)}$. Therefore, $\Lambda(\varphi_{02}(\beta))$ is
diagonal on $\{\Pi_i\}$ by noticing that
$[\Lambda(\varphi_{02}(\beta)),\Lambda(\phi_1)]=0$. Since the
channel cannot increase the distance between states,
$\Lambda(\varphi_{02}(\beta))$ breaks the same degeneracy as
$\Lambda(\phi_{0})$ for sufficiently large $|\beta|$. From Eq.
(\ref{degenerate}), we have $\Lambda(\varphi(\beta))$ and
$\Lambda(\varphi(-\beta))$ are diagonal on $\{\Pi_i\}$. Therefore,
$\Lambda(\varphi_{01})=\Lambda(\varphi(\beta))+\Lambda(\varphi(-\beta))-|\beta|^2\Lambda(\phi_2)$
is also diagonal on $\{\Pi_i\}$. Further, we will show that
$\Lambda(\varphi(\beta))$ is diagonal on $\{\Pi_i\}$ for arbitrary
$\beta$. From Eq. (\ref{degenerate}), this is obvious when
$\Lambda(\varphi_{02}(\beta))$ is nondegenerate. For the case where
$\Lambda(\varphi_{02}(\beta))$ is degenerate,
$\Lambda(\varphi_{02}(-\beta))=|\beta|^2\Lambda(\phi_0)+\Lambda(\phi_2)-\Lambda(\varphi_{02}(\beta))$
is nondegenerate and consequently, $\Lambda(\varphi(-\beta))$ is
diagonal on $\{\Pi_i\}$. Therefore,
$\Lambda(\varphi(\beta))=\Lambda(\varphi_{01})+|\beta|^2\Lambda(\phi_2)-\Lambda(\varphi(-\beta))$
is diagonal on $\{\Pi_i\}$. $\Lambda$ is a completely decohering
channel.

Case 3: now we are only left with the case that
$N^{\phi_0^{\bot}}=N^{\phi_0}$ but $\mathrm
I_{r_i(\phi_0^{\bot})}\neq\mathrm I_{r_i(\phi_0)}$, which can happen
only when $\Lambda(\mathrm I)=\mathrm I$ and $N^{\phi_k}=2$.
Therefore, we have
\begin{equation}
\Lambda(\phi_k)=p\Pi(\phi_k)+(1-p)\frac{\mathrm I}{3},
\end{equation}
where $\Pi(\phi_k)$ is a basis determined by $|\phi_k\rangle$.
Notices that $p$ is independent of $|\phi_k\rangle$ because of the
linearity of $\Lambda$. Consequently, for any input state
$\rho=\sum_ip_i|\alpha_i\rangle\langle\alpha_i|$, we have
$\Lambda(\rho)=p\sum_ip_i\Pi(\alpha_i)+(1-p)\mathrm I/3$. It means
that channel $\Lambda$ is an isotropic channel.

Combining the three cases together, we conclude that for a qutrit, a
commutativity-preserving channel is either a completely decohering
channel or an isotropic channel.

Since depolarizing channel is a subset of mixing channel, there
exist mixing channels that are able to locally create quantum
correlation. Therefore, mixedness can contribute to creation of
quantum correlations. Here we give an example to look more closely
at why a mixing channel can create quantum correlation in states
with high dimensions. Consider the following mixing channel
$\Lambda(\cdot)=\sum_i\mathrm E^{(i)}(\cdot)\mathrm E^{(i)\dagger}$,
where the Kraus operators are
\begin{eqnarray}
\mathrm E^{(0)}&=&|2\rangle\langle2|,\nonumber\\
\mathrm
E^{(i)}&=&e_iu_2^{(i)}(|0\rangle\langle0|+|1\rangle\langle1|),i=1,2,\cdots.
\end{eqnarray}
Here $u_2^{(i)}$ are rank-2 unitary operators on basis
$\{|0\rangle,|1\rangle\}$. This channel can create quantum
correlation in the state
$\rho=\tilde{\rho}_A^{\phi}\otimes|\phi\rangle_B\langle\phi|+\tilde{\rho}_A^{\psi}\otimes|\psi\rangle_B\langle\psi|$
if and only if Eq. (\ref{c3}) is violated. Writing the two
orthogonal states as $|\phi\rangle=\sum_{i=0}^2a_i|i\rangle$ and
$|\psi\rangle=\sum_{i=0}^2b_i|i\rangle$ ($\sum_{i=0}^2a_ib^*_i=0$),
we obtain the left hand side of Eq. (\ref{c3})
\begin{equation}
[\sum_ie_i^2u_2^{(i)}|\phi_2\rangle\langle\phi_2|u_2^{(i)\dagger},\sum_ie_i^2u_2^{(i)}|\psi_2\rangle\langle\psi_2|u_2^{(i)\dagger}],
\end{equation}
where $|\phi_2\rangle=a_0|0\rangle+a_1|0\rangle$ and
$|\psi_2\rangle=b_0|0\rangle+b_1|0\rangle$ are reduced states on
Hilbert space of dimension 2. Therefore, Eq. (\ref{c3}) is violated
if and only if $\langle\phi_2|\psi_2\rangle\neq0,1$. Two
high-dimension orthogonal states may become unorthogonal when
reduced to Hilbert space of dimension two. This is just the reason
for creating quantum correlation using a local mixing channel.
Isotropic channels act on all of the states in Hilbert spaces
equivalently, so they are likely the only subset of mixing channels
which belongs to the class of commutativity-preserving channels.
This observation leads to the following conjecture.

Conjecture: Consider the half-classical input state in Eq.
(\ref{half_classical}) where $B$ is a $d$-dimension quantum system
(qudit) with $d\geq3$, a channel $\Lambda$ can not create quantum
correlation in any half-classical input state if and only if
$\Lambda$ is either a completely dechering channel or an isotropic
channel.

We further prove that mixing channel can not increase the
teleportation fidelity of any two-qudit state. The average
teleportation fidelity $f$ is related to the maximum singlet
fraction (MSF) \cite{Horodecki1999}
$F=\max_{\Phi}\langle\Phi|\rho|\Phi\rangle$ as $f=(dF+1)/(d+1)$.
After the action of mixing channel on $B$, the MSF becomes
\begin{equation}
F'=\mathrm{Tr}(\rho\Xi)
\end{equation}
where $\Xi=\sum_i\mathrm I\otimes\mathrm
E^{(i)\dagger}|\Phi\rangle\langle\Phi|\mathrm I\otimes\mathrm
E^{(i)}$. Notice that for a mixing channel
$\Lambda(\cdot)=\sum_i\mathrm E^{(i)}(\cdot)\mathrm E^{(i)\dagger}$,
its conjecture $\Lambda^*(\cdot)=\sum_i\mathrm
E^{(i)\dagger}(\cdot)\mathrm E^{(i)}$ is also a mixing channel.
Therefore, $\Xi_A=\Xi_B=\mathrm I/2$, so $\Xi$ can be decomposed as
a mixture of maximally entangled pure states
$\Xi=\sum_ip_i|\Phi_i\rangle\langle\Phi_i|$. Then we have
$F'=\sum_ip_i\langle\Phi_i|\rho|\Phi_i\rangle\leq F$. Therefore,
average teleportation fidelity can never be increased by mixing
channel. This result suggests that quantum correlation created by
mixing channel may not be a useful resource for quantum information
tasks.

In summary, we have proved that the necessary and sufficient
condition for a local operation to create quantum correlation in
some half-classical state is that it is not a
commutativity-preserving channel. When the subsystem $B$ affected by
the local channel is a qubit, a commutativity-preserving channel is
either a mixing channel or a completely decohering channel. This
result confirms the results in Ref. \cite{PhysRevLett.107.170502}.
When $B$ is a qutrit, we have proved that a commutativity-preserving
channel is either an isotropic channel or a completely decohering
channel. This result is likely to be extended to arbitrary finite
dimension situation.

Hu thanks Sixia Yu and Chengjie Zhang for helpful discussions. This
work is supported NSFC under grants Nos. 10934010, 60978019, the
NKBRSFC under grants Nos. 2009CB930701, 2010CB922904, 2011CB921502,
2012CB821300, NSFC-RGC under grants Nos. 11061160490,
1386-N-HKU748/10, and CNSF under grants Nos. 10975181 and 11175247.

\newpage 
\bibliography{apssamp}

\begin{thebibliography}{28}
\expandafter\ifx\csname natexlab\endcsname\relax\def\natexlab#1{#1}\fi
\expandafter\ifx\csname bibnamefont\endcsname\relax
  \def\bibnamefont#1{#1}\fi
\expandafter\ifx\csname bibfnamefont\endcsname\relax
  \def\bibfnamefont#1{#1}\fi
\expandafter\ifx\csname citenamefont\endcsname\relax
  \def\citenamefont#1{#1}\fi
\expandafter\ifx\csname url\endcsname\relax
  \def\url#1{\texttt{#1}}\fi
\expandafter\ifx\csname urlprefix\endcsname\relax\def\urlprefix{URL }\fi
\providecommand{\bibinfo}[2]{#2}
\providecommand{\eprint}[2][]{\url{#2}}

\bibitem[{\citenamefont{Bennett et~al.}(1993)\citenamefont{Bennett, Brassard,
  Cr\'epeau, Jozsa, Peres, and Wootters}}]{tele}
\bibinfo{author}{\bibfnamefont{C.~H.} \bibnamefont{Bennett}},
  \bibinfo{author}{\bibfnamefont{G.}~\bibnamefont{Brassard}},
  \bibinfo{author}{\bibfnamefont{C.}~\bibnamefont{Cr\'epeau}},
  \bibinfo{author}{\bibfnamefont{R.}~\bibnamefont{Jozsa}},
  \bibinfo{author}{\bibfnamefont{A.}~\bibnamefont{Peres}}, \bibnamefont{and}
  \bibinfo{author}{\bibfnamefont{W.~K.} \bibnamefont{Wootters}},
  \bibinfo{journal}{Phys.\ Rev.\ Lett.} \textbf{\bibinfo{volume}{70}},
  \bibinfo{pages}{1895} (\bibinfo{year}{1993}).

\bibitem[{\citenamefont{Hausladen et~al.}(1996)\citenamefont{Hausladen, Jozsa,
  Schumacher, Westmoreland, and Wootters}}]{PhysRevA.54.1869}
\bibinfo{author}{\bibfnamefont{P.}~\bibnamefont{Hausladen}},
  \bibinfo{author}{\bibfnamefont{R.}~\bibnamefont{Jozsa}},
  \bibinfo{author}{\bibfnamefont{B.}~\bibnamefont{Schumacher}},
  \bibinfo{author}{\bibfnamefont{M.}~\bibnamefont{Westmoreland}},
  \bibnamefont{and} \bibinfo{author}{\bibfnamefont{W.~K.}
  \bibnamefont{Wootters}}, \bibinfo{journal}{Phys. Rev. A}
  \textbf{\bibinfo{volume}{54}}, \bibinfo{pages}{1869} (\bibinfo{year}{1996}).

\bibitem[{\citenamefont{Ollivier and Zurek}(2001)}]{PhysRevLett.88.017901}
\bibinfo{author}{\bibfnamefont{H.}~\bibnamefont{Ollivier}} \bibnamefont{and}
  \bibinfo{author}{\bibfnamefont{W.~H.} \bibnamefont{Zurek}},
  \bibinfo{journal}{Phys. Rev. Lett.} \textbf{\bibinfo{volume}{88}},
  \bibinfo{pages}{017901} (\bibinfo{year}{2001}).

\bibitem[{\citenamefont{Oppenheim et~al.}(2002)\citenamefont{Oppenheim,
  Horodecki, Horodecki, and Horodecki}}]{PhysRevLett.89.180402}
\bibinfo{author}{\bibfnamefont{J.}~\bibnamefont{Oppenheim}},
  \bibinfo{author}{\bibfnamefont{M.}~\bibnamefont{Horodecki}},
  \bibinfo{author}{\bibfnamefont{P.}~\bibnamefont{Horodecki}},
  \bibnamefont{and}
  \bibinfo{author}{\bibfnamefont{R.}~\bibnamefont{Horodecki}},
  \bibinfo{journal}{Phys. Rev. Lett.} \textbf{\bibinfo{volume}{89}},
  \bibinfo{pages}{180402} (\bibinfo{year}{2002}).

\bibitem[{\citenamefont{Horodecki et~al.}(2005)\citenamefont{Horodecki,
  Horodecki, Horodecki, Oppenheim, Sen(De), Sen, and
  Synak-Radtke}}]{PhysRevA.71.062307}
\bibinfo{author}{\bibfnamefont{M.}~\bibnamefont{Horodecki}},
  \bibinfo{author}{\bibfnamefont{P.}~\bibnamefont{Horodecki}},
  \bibinfo{author}{\bibfnamefont{R.}~\bibnamefont{Horodecki}},
  \bibinfo{author}{\bibfnamefont{J.}~\bibnamefont{Oppenheim}},
  \bibinfo{author}{\bibfnamefont{A.}~\bibnamefont{Sen(De)}},
  \bibinfo{author}{\bibfnamefont{U.}~\bibnamefont{Sen}}, \bibnamefont{and}
  \bibinfo{author}{\bibfnamefont{B.}~\bibnamefont{Synak-Radtke}},
  \bibinfo{journal}{Phys. Rev. A} \textbf{\bibinfo{volume}{71}},
  \bibinfo{pages}{062307} (\bibinfo{year}{2005}).

\bibitem[{\citenamefont{Datta et~al.}(2008)\citenamefont{Datta, Shaji, and
  Caves}}]{PhysRevLett.100.050502}
\bibinfo{author}{\bibfnamefont{A.}~\bibnamefont{Datta}},
  \bibinfo{author}{\bibfnamefont{A.}~\bibnamefont{Shaji}}, \bibnamefont{and}
  \bibinfo{author}{\bibfnamefont{C.~M.} \bibnamefont{Caves}},
  \bibinfo{journal}{Phys. Rev. Lett.} \textbf{\bibinfo{volume}{100}},
  \bibinfo{pages}{050502} (\bibinfo{year}{2008}).

\bibitem[{\citenamefont{Roa et~al.}(2011)\citenamefont{Roa, Retamal, and
  Alid-Vaccarezza}}]{PhysRevLett.107.080401}
\bibinfo{author}{\bibfnamefont{L.}~\bibnamefont{Roa}},
  \bibinfo{author}{\bibfnamefont{J.~C.} \bibnamefont{Retamal}},
  \bibnamefont{and}
  \bibinfo{author}{\bibfnamefont{M.}~\bibnamefont{Alid-Vaccarezza}},
  \bibinfo{journal}{Phys. Rev. Lett.} \textbf{\bibinfo{volume}{107}},
  \bibinfo{pages}{080401} (\bibinfo{year}{2011}).

\bibitem[{\citenamefont{Luo}(2008)}]{PhysRevA.77.042303}
\bibinfo{author}{\bibfnamefont{S.}~\bibnamefont{Luo}}, \bibinfo{journal}{Phys.
  Rev. A} \textbf{\bibinfo{volume}{77}}, \bibinfo{pages}{042303}
  (\bibinfo{year}{2008}).

\bibitem[{\citenamefont{Ali et~al.}(2010)\citenamefont{Ali, Rau, and
  Alber}}]{PhysRevA.81.042105}
\bibinfo{author}{\bibfnamefont{M.}~\bibnamefont{Ali}},
  \bibinfo{author}{\bibfnamefont{A.~R.~P.} \bibnamefont{Rau}},
  \bibnamefont{and} \bibinfo{author}{\bibfnamefont{G.}~\bibnamefont{Alber}},
  \bibinfo{journal}{Phys. Rev. A} \textbf{\bibinfo{volume}{81}},
  \bibinfo{pages}{042105} (\bibinfo{year}{2010}).

\bibitem[{\citenamefont{Coles}(2011)}]{arXiv:1110.1664v1}
\bibinfo{author}{\bibfnamefont{P.~J.} \bibnamefont{Coles}},
  \bibinfo{journal}{arXiv: 1110.1664v1}  (\bibinfo{year}{2011}).

\bibitem[{\citenamefont{Maruyama et~al.}(2009)\citenamefont{Maruyama, Nori, and
  Vedral}}]{RevModPhys.81.1}
\bibinfo{author}{\bibfnamefont{K.}~\bibnamefont{Maruyama}},
  \bibinfo{author}{\bibfnamefont{F.}~\bibnamefont{Nori}}, \bibnamefont{and}
  \bibinfo{author}{\bibfnamefont{V.}~\bibnamefont{Vedral}},
  \bibinfo{journal}{Rev. Mod. Phys.} \textbf{\bibinfo{volume}{81}},
  \bibinfo{pages}{1} (\bibinfo{year}{2009}).

\bibitem[{\citenamefont{Zurek}(2003)}]{PhysRevA.67.012320}
\bibinfo{author}{\bibfnamefont{W.~H.} \bibnamefont{Zurek}},
  \bibinfo{journal}{Phys. Rev. A} \textbf{\bibinfo{volume}{67}},
  \bibinfo{pages}{012320} (\bibinfo{year}{2003}).

\bibitem[{\citenamefont{Cornelio et~al.}(2011)\citenamefont{Cornelio,
  de~Oliveira, and Fanchini}}]{PhysRevLett.107.020502}
\bibinfo{author}{\bibfnamefont{M.~F.} \bibnamefont{Cornelio}},
  \bibinfo{author}{\bibfnamefont{M.~C.} \bibnamefont{de~Oliveira}},
  \bibnamefont{and} \bibinfo{author}{\bibfnamefont{F.~F.}
  \bibnamefont{Fanchini}}, \bibinfo{journal}{Phys. Rev. Lett.}
  \textbf{\bibinfo{volume}{107}}, \bibinfo{pages}{020502}
  (\bibinfo{year}{2011}).

\bibitem[{\citenamefont{Streltsov
  et~al.}(2011{\natexlab{a}})\citenamefont{Streltsov, Kampermann, and
  Bru\ss{}}}]{PhysRevLett.106.160401}
\bibinfo{author}{\bibfnamefont{A.}~\bibnamefont{Streltsov}},
  \bibinfo{author}{\bibfnamefont{H.}~\bibnamefont{Kampermann}},
  \bibnamefont{and} \bibinfo{author}{\bibfnamefont{D.}~\bibnamefont{Bru\ss{}}},
  \bibinfo{journal}{Phys. Rev. Lett.} \textbf{\bibinfo{volume}{106}},
  \bibinfo{pages}{160401} (\bibinfo{year}{2011}{\natexlab{a}}).

\bibitem[{\citenamefont{Piani et~al.}(2011)\citenamefont{Piani, Gharibian,
  Adesso, Calsamiglia, Horodecki, and Winter}}]{PhysRevLett.106.220403}
\bibinfo{author}{\bibfnamefont{M.}~\bibnamefont{Piani}},
  \bibinfo{author}{\bibfnamefont{S.}~\bibnamefont{Gharibian}},
  \bibinfo{author}{\bibfnamefont{G.}~\bibnamefont{Adesso}},
  \bibinfo{author}{\bibfnamefont{J.}~\bibnamefont{Calsamiglia}},
  \bibinfo{author}{\bibfnamefont{P.}~\bibnamefont{Horodecki}},
  \bibnamefont{and} \bibinfo{author}{\bibfnamefont{A.}~\bibnamefont{Winter}},
  \bibinfo{journal}{Phys. Rev. Lett.} \textbf{\bibinfo{volume}{106}},
  \bibinfo{pages}{220403} (\bibinfo{year}{2011}).

\bibitem[{\citenamefont{Badzi\c{a}g et~al.}(2000)\citenamefont{Badzi\c{a}g,
  Horodecki, Horodecki, and Horodecki}}]{PhysRevA.62.012311}
\bibinfo{author}{\bibfnamefont{P.}~\bibnamefont{Badzi\c{a}g}},
  \bibinfo{author}{\bibfnamefont{M.}~\bibnamefont{Horodecki}},
  \bibinfo{author}{\bibfnamefont{P.}~\bibnamefont{Horodecki}},
  \bibnamefont{and}
  \bibinfo{author}{\bibfnamefont{R.}~\bibnamefont{Horodecki}},
  \bibinfo{journal}{Phys. Rev. A} \textbf{\bibinfo{volume}{62}},
  \bibinfo{pages}{012311} (\bibinfo{year}{2000}).

\bibitem[{\citenamefont{Yeo}(2008)}]{PhysRevA.78.022334}
\bibinfo{author}{\bibfnamefont{Y.}~\bibnamefont{Yeo}}, \bibinfo{journal}{Phys.
  Rev. A} \textbf{\bibinfo{volume}{78}}, \bibinfo{pages}{022334}
  (\bibinfo{year}{2008}).

\bibitem[{\citenamefont{Hu et~al.}(2010)\citenamefont{Hu, Gu, Gong, and
  Guo}}]{PhysRevA.81.054302}
\bibinfo{author}{\bibfnamefont{X.}~\bibnamefont{Hu}},
  \bibinfo{author}{\bibfnamefont{Y.}~\bibnamefont{Gu}},
  \bibinfo{author}{\bibfnamefont{Q.}~\bibnamefont{Gong}}, \bibnamefont{and}
  \bibinfo{author}{\bibfnamefont{G.}~\bibnamefont{Guo}},
  \bibinfo{journal}{Phys. Rev. A} \textbf{\bibinfo{volume}{81}},
  \bibinfo{pages}{054302} (\bibinfo{year}{2010}).

\bibitem[{\citenamefont{Hu et~al.}(2011)\citenamefont{Hu, Gu, Gong, and
  Guo}}]{PhysRevA.84.022113}
\bibinfo{author}{\bibfnamefont{X.}~\bibnamefont{Hu}},
  \bibinfo{author}{\bibfnamefont{Y.}~\bibnamefont{Gu}},
  \bibinfo{author}{\bibfnamefont{Q.}~\bibnamefont{Gong}}, \bibnamefont{and}
  \bibinfo{author}{\bibfnamefont{G.}~\bibnamefont{Guo}},
  \bibinfo{journal}{Phys. Rev. A} \textbf{\bibinfo{volume}{84}},
  \bibinfo{pages}{022113} (\bibinfo{year}{2011}).

\bibitem[{\citenamefont{Ciccarello and
  Giovannetti}(2012{\natexlab{a}})}]{PhysRevA.85.010102}
\bibinfo{author}{\bibfnamefont{F.}~\bibnamefont{Ciccarello}} \bibnamefont{and}
  \bibinfo{author}{\bibfnamefont{V.}~\bibnamefont{Giovannetti}},
  \bibinfo{journal}{Phys. Rev. A} \textbf{\bibinfo{volume}{85}},
  \bibinfo{pages}{010102} (\bibinfo{year}{2012}{\natexlab{a}}).

\bibitem[{\citenamefont{Ciccarello and
  Giovannetti}(2012{\natexlab{b}})}]{PhysRevA.85.022108}
\bibinfo{author}{\bibfnamefont{F.}~\bibnamefont{Ciccarello}} \bibnamefont{and}
  \bibinfo{author}{\bibfnamefont{V.}~\bibnamefont{Giovannetti}},
  \bibinfo{journal}{Phys. Rev. A} \textbf{\bibinfo{volume}{85}},
  \bibinfo{pages}{022108} (\bibinfo{year}{2012}{\natexlab{b}}).

\bibitem[{\citenamefont{Li and Luo}(2008)}]{PhysRevA.78.024303}
\bibinfo{author}{\bibfnamefont{N.}~\bibnamefont{Li}} \bibnamefont{and}
  \bibinfo{author}{\bibfnamefont{S.}~\bibnamefont{Luo}},
  \bibinfo{journal}{Phys. Rev. A} \textbf{\bibinfo{volume}{78}},
  \bibinfo{pages}{024303} (\bibinfo{year}{2008}).

\bibitem[{\citenamefont{Ferraro et~al.}(2010)\citenamefont{Ferraro, Aolita,
  Cavalcanti, Cucchietti, and Ac\'\i{}n}}]{PhysRevA.81.052318}
\bibinfo{author}{\bibfnamefont{A.}~\bibnamefont{Ferraro}},
  \bibinfo{author}{\bibfnamefont{L.}~\bibnamefont{Aolita}},
  \bibinfo{author}{\bibfnamefont{D.}~\bibnamefont{Cavalcanti}},
  \bibinfo{author}{\bibfnamefont{F.~M.} \bibnamefont{Cucchietti}},
  \bibnamefont{and}
  \bibinfo{author}{\bibfnamefont{A.}~\bibnamefont{Ac\'\i{}n}},
  \bibinfo{journal}{Phys. Rev. A} \textbf{\bibinfo{volume}{81}},
  \bibinfo{pages}{052318} (\bibinfo{year}{2010}).

\bibitem[{\citenamefont{Streltsov
  et~al.}(2011{\natexlab{b}})\citenamefont{Streltsov, Kampermann, and
  Bru\ss{}}}]{PhysRevLett.107.170502}
\bibinfo{author}{\bibfnamefont{A.}~\bibnamefont{Streltsov}},
  \bibinfo{author}{\bibfnamefont{H.}~\bibnamefont{Kampermann}},
  \bibnamefont{and} \bibinfo{author}{\bibfnamefont{D.}~\bibnamefont{Bru\ss{}}},
  \bibinfo{journal}{Phys. Rev. Lett.} \textbf{\bibinfo{volume}{107}},
  \bibinfo{pages}{170502} (\bibinfo{year}{2011}{\natexlab{b}}).

\bibitem[{\citenamefont{Datta}(2008)}]{arXiv:0807.4490v1}
\bibinfo{author}{\bibfnamefont{A.}~\bibnamefont{Datta}},
  \bibinfo{journal}{arXiv:0807.4490v1}  (\bibinfo{year}{2008}).

\bibitem[{\citenamefont{Wehrl}(1978)}]{RevModPhys.50.221}
\bibinfo{author}{\bibfnamefont{A.}~\bibnamefont{Wehrl}}, \bibinfo{journal}{Rev.
  Mod. Phys.} \textbf{\bibinfo{volume}{50}}, \bibinfo{pages}{221}
  (\bibinfo{year}{1978}).

\bibitem[{\citenamefont{Marcus and Moyls}(1959)}]{EnPC}
\bibinfo{author}{\bibfnamefont{M.}~\bibnamefont{Marcus}} \bibnamefont{and}
  \bibinfo{author}{\bibfnamefont{B.~N.} \bibnamefont{Moyls}},
  \bibinfo{journal}{Canad. J. Math.} \textbf{\bibinfo{volume}{11}},
  \bibinfo{pages}{61} (\bibinfo{year}{1959}).

\bibitem[{\citenamefont{Horodecki et~al.}(1999)\citenamefont{Horodecki,
  Horodecki, and Horodecki}}]{Horodecki1999}
\bibinfo{author}{\bibfnamefont{M.}~\bibnamefont{Horodecki}},
  \bibinfo{author}{\bibfnamefont{P.}~\bibnamefont{Horodecki}},
  \bibnamefont{and}
  \bibinfo{author}{\bibfnamefont{R.}~\bibnamefont{Horodecki}},
  \bibinfo{journal}{Phys.\ Rev.\ A} \textbf{\bibinfo{volume}{60}},
  \bibinfo{pages}{1888} (\bibinfo{year}{1999}).

\end{thebibliography}

\end{document}